\documentclass[letterpaper, 10 pt, conference]{ieeeconf}  

\IEEEoverridecommandlockouts                              

\usepackage{subcaption}
\usepackage[utf8]{inputenc}
\PassOptionsToPackage{dvipsnames, table}{xcolor}
\usepackage{color}
\usepackage{array}
\usepackage{verbatim}
\usepackage{float}
\usepackage{multirow}
\usepackage{amsmath,bm}
\usepackage{amssymb}
\usepackage{hyperref}
\usepackage{tabularx}
\usepackage{slashed}
\usepackage{svg}
\usepackage{ifsym}
\usepackage{wasysym}
\usepackage{tikz}
\usepackage{pgfplots}
\pgfplotsset{compat=1.17}
\usepackage{caption}
\usepackage{wrapfig}
\usepackage{diagbox}
\usepackage{booktabs}
\usepackage{cite}
\usepackage{url}
\usepackage{color}
\usepackage[dvipsnames]{xcolor}

\DeclareMathOperator*{\argmax}{arg\,max}
\title{\LARGE \bf
Uncertainty-Aware Guidance for Target Tracking subject to\\ Intermittent Measurements using Motion Model Learning}

\author{Andres Pulido$^{1}$, Kyle Volle$^{2}$, Kristy Waters$^{1}$, Zachary I. Bell$^{3}$, Prashant Ganesh$^{4}$ and Jane Shin$^{1}$
\thanks{$^{1}$Andres Pulido, Kristy Waters, and Jane Shin are with the department of Mechanical and Aerospace Engineering,
        University of Florida, Gainesville, FL 32611
        {\tt\small \{andrespulido, watersk, jane.shin\}@ufl.edu}}%
\thanks{$^{2}$Kyle Volle was previously affiliated with University of Florida and is now with Torch Technologies,
        6 11th Ave Suite F-1, Shalimar, FL 32579
        {\tt\small kvolle@torchtechnologies.com}}%
\thanks{$^{3}$Zachary I. Bell is with Air Force Research Lab,
        203 Eglin Blvd, Eglin AFB, FL 32542
        {\tt\small zachary.bell.10@us.af.mil}}%
\thanks{$^{4}$Prashant Ganesh was previously affiliated with University of Florida and now is with EpiSci,
        13025 Danielson St Ste \#106, Poway, CA 92064
        {\tt\small prashantganesh@episci.com}}%
}

\begin{document}

\maketitle
\thispagestyle{empty}
\pagestyle{empty}

\begin{abstract}
This paper presents a novel guidance law for target tracking applications where the target motion model is unknown and sensor measurements are intermittent due to unknown environmental conditions and low measurement update rate. In this work, the target motion model is represented by a transformer neural network and trained by previous target position measurements. This transformer motion model serves as the prediction step in a particle filter for target state estimation and uncertainty quantification. The particle filter estimation uncertainty is utilized in the information-driven guidance law to compute a path for the mobile agent to travel to a position with maximum expected entropy reduction (EER). The computation of EER is performed in real-time by approximating the information gain from the predicted particle distributions relative to the current distribution. Simulation and hardware experiments are performed with a quadcopter agent and TurtleBot target to demonstrate that the presented guidance law outperforms two other baseline guidance methods.
\end{abstract}

\section{INTRODUCTION}

\begin{figure}[t!]
    \centering
    \includegraphics[width=0.45\textwidth]{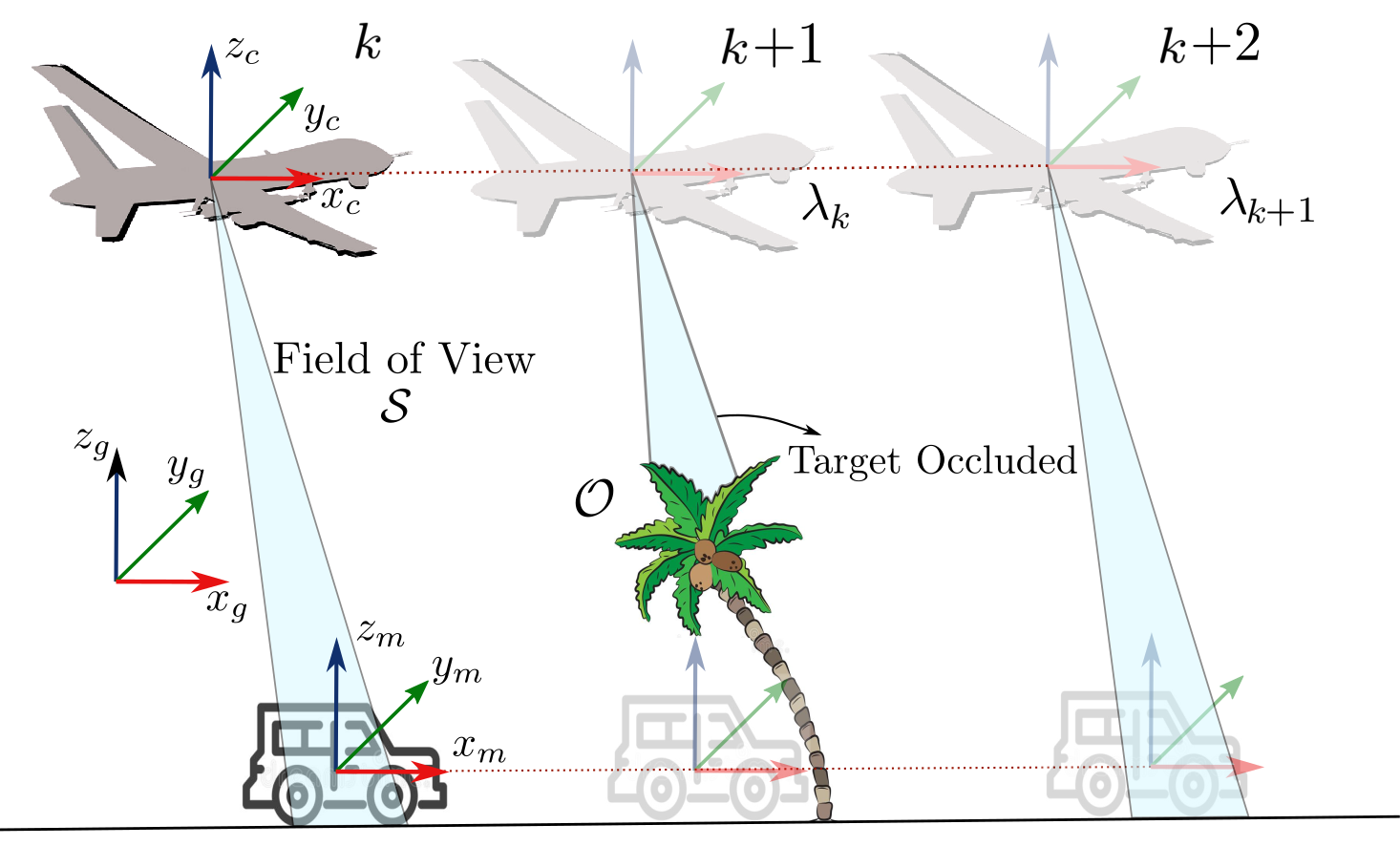}
    \caption{An example scenario of target tracking with intermittent measurements.}
    \label{fig:occlusion_example}
\end{figure}

Target tracking refers to a capability of an agent to estimate a motion model or dynamics of a moving target based on sensor measurements. Target tracking problems arise in many robotics applications such as drone search and rescue, and the problem becomes more challenging as sensor measurements become poor or unavailable intermittently, possibly due to occlusion \cite{rosencrantz2003locating}. For example, as shown in Figure \ref{fig:occlusion_example}, when a drone (agent) tracks a ground vehicle (target) for reconnaissance, the target may be occluded by obstacles, such as buildings or vegetation, from the agent's sensor field-of-view (FOV). When the measurements are unavailable, the agent can only predict the target position using approximated motion models. Therefore, the agent must plan its path considering the estimated motion model so that the agent can obtain target measurements once the target becomes visible and reduce the uncertainty associated with target estimate.

To improve the time the target is being observed, it is essential to obtain an accurate motion model to predict the target state when sensor measurements are unavailable. The agent can utilize prior observations of the target to train motion models offline; however, recent work has demonstrated a real-time method to train a motion model network \cite{parikhTargetTrackingPresence2018} and attention-based deep motion model networks (DMMN) \cite{bellTargetTrackingSubject2023} based on the intermittent camera measurements obtained online. In \cite{harrisTargetTrackingPresence2020} and \cite{mccourtPassivityBasedTargetTracking2022}, the authors study dwell-time conditions to find the time that the target can leave the sensor FOV while having a bounded estimation error. However, the agent’s guidance approach, which plans the agent’s next waypoint to observe the target and reduce the estimation uncertainty, has not been studied in the aforementioned related research that utilize learned motion models.

Information-theoretic approaches have proven effective in planning the agent’s next waypoint when onboard sensors need to obtain certain measurements to reduce estimation uncertainty \cite{luComparisonInformationTheoretic2012, shinInformativeMultiviewPlanning2022a}. These information-driven planning approaches find the best action, or waypoint, for the agent that gives the highest expected information gain with respect to the measurements at a future time step. Additionally, due to its high computational complexity, particle filters have been used to approximate the probability distributions used in the computation \cite{boersParticleFilterBased2010, thrun2005probabilistic, rui_path_2018, skoglar_information_2009}. These information-driven planning approaches implemented with particle filters have been proven to be effective in other applications, including simultaneous localization and mapping (SLAM) \cite{StachnissInfoGainExplorationRBPF05}. Another information-based tracking approach is demonstrated in \cite{ramirez-paredesDistributedInformationbasedGuidance2018}, where the authors use optimal information gain to track a target with multiple agents; however, \cite{ramirez-paredesDistributedInformationbasedGuidance2018} assumes that the road network, along which the targets move, is known so that the cost function can use this knowledge on the road network to find the waypoints to maximize the information gain and avoid collisions with other agents.

In this paper, a novel real-time, image-based uncertainty-aware guidance law is proposed to track a moving target, with unknown dynamics model, subject to occlusions caused by unknown environmental conditions and physical limitations of the camera FOV (Problem Statement in Section \ref{sec:problem_formulation}). This novel approach utilizes a transformer-based DMMN in a particle filter estimator to predict the state of a moving target, whose uncertainty is then utilized by a following novel information-driven planning algorithm that computes the path for the mobile agent to reduce the target's localization uncertainty. This paper uses expected entropy reduction (EER) as an expected information gain in order to directly account for target's localization uncertainty. A sampling method is developed to estimate the EER using a subset of the particles and for the DMMN to enable real-time hardware demonstration of the framework (described in Sections \ref{sec:method} and \ref{sec:setup}). This proposed uncertainty-aware guidance law for target tracking is then compared to other baseline approaches (Section \ref{sec:results}). The code and data for this paper are open-source\footnote{\url{https://github.com/aprilab-uf/info_driven_guidance}}.

\section{PROBLEM STATEMENT}
\label{sec:problem_formulation}
This work considers the problem of finding a guidance law of an aerial agent that is tasked with tracking a moving target on the ground. The agent is equipped with a fixed downward-facing camera that uses images to measure the position the target. The agent is assumed to operate at a constant height such that the projected geometry of the camera's field-of-view (FOV) remains constant. The target moves along a road network that is unknown to the agent and which is learned using the DMMN from measurements available \textit{a priori}. To achieve the tracking objective, the agent uses the DMMN-based particle filter to estimate the target's current position and predict the future position at any time. Using these predictions, the agent approximates the guidance law that will minimize the DMMN-based particle filter future uncertainty.

\begin{figure}[!htbp]
    \centering
    \smallskip
    \smallskip
    \includegraphics[width=0.40\textwidth]{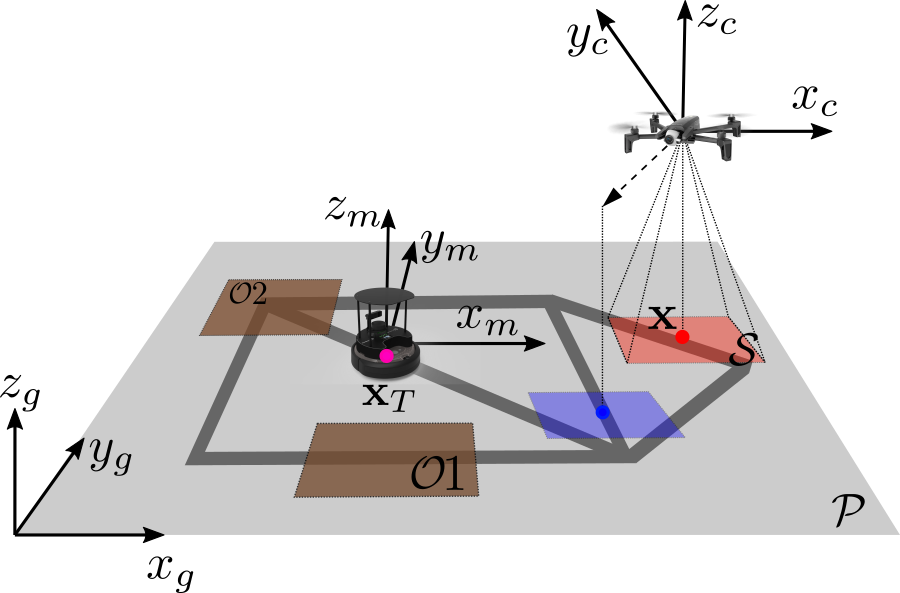}
    \caption{A schematic of the workspace and coordinates}
    \label{fig:workspace}
\end{figure}

The workspace is defined as a 3D Euclidean space, $\mathcal{W}\subset \mathbb{R}^{3}$, in which three coordinate frames are defined as shown in Figure \ref{fig:workspace}. The inertial frame is denoted by $\mathcal{F}_{g}$. 
The agent's camera frame, $\mathcal{F}_{c}$, is defined with the origin at the principal point of the camera, denoted by $c$. The mobile target frame, denoted by $\mathcal{F}_{m}$, has an origin located at the center of the target, denoted as $m$. The target is only moving along the road network on $x_{g}y_{g}$-plane, denoted by $\mathcal{P} \subset \mathbb{R}^{2}$. The projected camera sensor FOV on is denoted by $\mathcal{S} \subset \mathcal{P}$.

Based on the aforementioned assumptions, the target state is defined by $\mathbf{x} = [x~ y]^{\top} \in \mathcal{P}$ with respect to $\mathcal{F}_{g}$. The center of projected sensor FOV $\mathcal{S}$, which is also the agent's position, is defined by $\mathbf{s} = [x_s~ y_s]^{\top} \in \mathbb{R}^{2}$. The unknown area where the target is occluded is denoted by $\mathcal{O}$, where $\mathcal{O}\subset \mathcal{P}$, i.e., the target is occluded when $\mathbf{x}\in\mathcal{O}$. This subset $\mathcal{O}$ is the union of compact occlusion regions denoted by $\mathcal{O}_{1}, ..., \mathcal{O}_{p}$, where $p$ is the number of occlusion regions. Then, the target will be observed if and only if $\mathbf{x} \in \mathcal{S} \cap \mathcal{O}^{c}$. When the target is observed, the agent obtains a measurement of the target state at discrete time step $k \in \mathbb{N}_{0}$, denoted by $\mathbf{z}_k = [\tilde{x}_{k}~ \tilde{y}_{k}]^{\top} \subset \mathcal{P}$, where $\tilde{x}_{k}$ and $\tilde{y}_{k}$ are the estimated target position in $\mathcal{F}_g$ at $k$. A state's value at time step $k$ is denoted by $(\cdot)_{k}$, and the set of state's value from time step $k_{1}$ to $k_{2}>k_{1}$ is denoted by $(\cdot)_{k_{1}:k_{2}}$.

The sensor measurement model $\textbf{h}$ is assumed to be probabilistic and known, denoted by
\begin{equation}
    \mathbf{z}_k = \mathbf{h}(\mathbf{x}_{k}, \mathbf{n}_{k})
    \label{eq:measurement_model}
\end{equation}
where $\mathbf{n}_{k}$ denotes a random noise from a known distribution. A set of measurements taken from time step $k_{1}$ to $k_{2}>k_{1}$ is denoted by $\mathbf{z}_{k_{1}:k_{2}}$. The unknown motion model is represented by
\begin{equation}
    \mathbf{x}_k = \mathbf{f}(\mathbf{x}_{k-1}, \mathbf{v})
    \label{eq:motion_model}
\end{equation}
where $\mathbf{v}$ is an unknown process noise. The objective of the guidance law is to find the future waypoint of an agent so that target tracking uncertainty, i.e., the uncertainty in target state estimation, is minimized.

\section{UNCERTAINTY-AWARE GUIDANCE WITH MOTION MODEL LEARNING}
\label{sec:method}

In order to learn the motion model of the target, the presented method uses a transformer-based DMMN (Section \ref{ssec:mml}), which is then inserted in target state estimation by approximating the probability distribution using particles (Section \ref{ssec:state_estimation}). Then, the proposed guidance law computes the next waypoint to reduce the target state estimation uncertainty using expected entropy reduction (Section \ref{ssec:eer_guidance}).

\subsection{Transformer-based Target Motion Model Learning}\label{ssec:mml}

The target's motion model is assumed unknown to the tracking agent; however, the motion model can be trained using previous measurements of the target. In this paper, the DMMN is trained offline with the input being a set of $K_{in}$ position histories $\mathbf{x}_{(k-K_{in}):k}$, for $k \in \{K_{in}, K_{in}+1, ...\}$ and output a future position of the target $\mathbf{x}_{k+K}$. By default, the network predicts a single time step into the future, but by appending the prediction to the input and feeding it through the network again, the tracking agent can predict the targets future position over an arbitrary time-horizon. 


In this paper, we design a DMMN transformer network based on \cite{ashishAttention} to leverage the transformer's ability to encode sequence of features. The transformer is comprised of a position encoder, a multi-head transformer encoder, and a linear decoder with dropout and ReLU activation functions. The encoder takes the time series data and maps them to points in a high-dimensional feature space. The decoder takes these points and maps them to a predicted position. This model is generalizable to target systems that operate under dynamics that can be approximated as a mapping between a trajectory and a predicted position.

\subsection{Target State Estimation using Particle Filter}
\label{ssec:state_estimation}
The target state can be recursively estimated using Bayesian inference. Specifically, the target state is estimated through two steps. The first step is called prediction, which computes
\begin{equation}
	p(\mathbf{x}_{k}~|~\mathbf{z}_{1:k-1}) = \int p(\mathbf{x}_{k}~|~\mathbf{x}_{k-1})p(\mathbf{x}_{k-1}~|~\mathbf{z}_{1:k-1})d\mathbf{x}_{k-1}
	\label{eq:prediction}
\end{equation}
where $p(\mathbf{x}_{k}~|~\mathbf{x}_{k-1})$ represents how the target state transitions, corresponding to the motion model in \eqref{eq:motion_model}. In this paper, a transformer-based DMMN is developed to compute this motion model as described in section \ref{ssec:mml}. The next step is called update, represented as
\begin{equation}
	p(\mathbf{x}_{k}~|~\mathbf{z}_{1:k}) \propto p(\mathbf{z}_{k}~|~\mathbf{x}_{k})p(\mathbf{x}_{k}~|~\mathbf{z}_{1:k-1})
	\label{eq:update}
\end{equation}
where the probability density function $p(\mathbf{z}_{k}~|~\mathbf{x}_{k})$ is computed using the measurement model in \eqref{eq:measurement_model}. It is assumed that $\mathbf{z}_k \sim \mathcal{N}(\mathbf{x}_{k}, \bf{\Sigma})$ and the covariance matrix $\bf{\Sigma} \in \mathbb{R}^{2 \times 2}$ are known. 

Considering the non-linearity of the target motion model, the target state is estimated using a particle filter. In the particle filter, the probability density functions on the target state is estimated using sampling techniques \cite{thrun2005probabilistic}. At time step $k-1$,
\begin{equation}
    p(\mathbf{x}_{k-1}~|~\mathbf{z}_{1:k-1})  \approx \sum_{i=1}^{N} w^{(i)}_{k-1} \delta(\mathbf{x}_{k-1} - \mathbf{x}^{(i)}_{k-1})
    \label{eq:pf}
\end{equation}
where $\delta{(\cdot)}$ is the Dirac delta function and $N$ is the number of particles. The weights are constrained by $\sum^N_{i=1} w^{(i)}_k = 1$. 

\subsection{EER-based Guidance Law}
\label{ssec:eer_guidance}
This paper proposes a guidance law for target tracking based on EER that can be computed in real-time onboard an aerial drone. Since entropy is a function of the probability density function of target state estimation, the presented work approximates entropy using sampling-based approximation represented in \eqref{eq:pf}. From \cite{boersParticleFilterBased2010}, the entropy of target state estimation can be approximated by
\begin{multline}
    H(p(\mathbf{x}_{k}~|~\mathbf{z}_{1:k})) \approx \log \left( \sum_{i=1}^N p(\mathbf{z}_{k} ~|~ \mathbf{x}_{k}^{(i)})w_{k-1}^{(i)} \right)\\
    - \sum_{i=1}^{N} \log \left(p(\mathbf{z}_k ~|~ \mathbf{x}_{k}^{(i)})(\sum_{j=1}^N p(\mathbf{x}_{k}^{(i)} ~|~ \mathbf{x}_{k-1}^{(j)})w_{k-1}^{(j)})\right)w_{k}^{(i)}
    \label{eq:entropy_particle_update}
\end{multline}
When target measurements are unavailable, based on \cite{skoglar_information_2009}, the entropy can be approximated by the particles of the prior distribution as
\begin{multline}
    H(p(\mathbf{x}_{k}~|~\mathbf{z}_{1:k-1}) \approx \\ 
    - \sum_{i=1}^{N}\log \left( \sum_{j=1}^{N} p(\mathbf{x}_{k}^{(i)} ~|~ \mathbf{x}_{k-1}^{(j)})w_{k-1}^{(j)} \right)w_{k}^{(i)}
    \label{eq:entropy_particle_prediction}
\end{multline}

In the guidance law, the expected value of entropy reduction should be computed as a function of the agent's future waypoint. Let us denote the agent's waypoint at time step $k+K$ as $\lambda_{k+K}$, where $K\in\mathbb{N}$ is a user-selected constant representing the time step horizon to plan the next waypoint. Also, denote $\mathbf{\hat{z}}_{k+K}$ as the measurement obtained at $\lambda_{k+K}$, and $\mathbf{\hat{x}}_{k+K}$ as the updated target state estimation. The information gain is set as the entropy reduction, which is represented as
\begin{multline}
    I(\mathbf{\hat{z}}_{k+K}, \lambda_{k+K}) \\ 
    = H(p(\mathbf{x}_{k}~|~\mathbf{z}_{1:k})) - H(p(\mathbf{\hat{x}}_{k+K}~|~\mathbf{z}_{1:k},\mathbf{\hat{z}}_{k+K},\lambda_{k+K}))
\end{multline}
Then, the expected information gain (or EER) is computed by integrating over all possible measurements $\hat{\mathbf{z}}_{k+K}$, which is represented by
\begin{multline}
    EER(\lambda_{k+K}) = \mathbb{E}_{\hat{\mathbf{z}}_{k+K}}[I(\mathbf{\hat{z}}_{k+K}, \lambda_{k+K})] \\
    = \int_{\mathbf{\hat{z}}_{k+K}} p(\mathbf{\hat{z}}_{k+K}~|~\lambda_{k+K}) I(\mathbf{\hat{z}}_{k+K}, \lambda_{k+K}) d\mathbf{\hat{z}}_{k+K}
    \label{eq:eer_def}
\end{multline}

Since the computational complexity of EER approximation grows exponentially to the number of particles $N$ in \eqref{eq:pf}, it becomes computationally expensive to run onboard a drone. Specifically, the entropy computation takes significant time due to exponential complexity. Therefore, the probability density function $p(\mathbf{x}_{k}~|~\mathbf{z}_{1:k+K})$ is approximated by sampling $N_{H} \leq N$ particles to reduce computation time. The number of particles sampled for entropy computation is denoted by $N_{H}$. These $N_{H}$ particles are uniformly sampled from $p(\mathbf{x}_k~|~\mathbf{z}_{1:k-1})$ to make sure the distribution keeps the same shape. Therefore, if $\mathbf{\hat{z}}_{k+K} \in \mathcal{S} \cap \mathcal{O}^{c}$ at $k+K$, then

\begin{multline}\label{eq:part_entropy}
    H(p(\mathbf{\hat{x}}_{k+K}~|~\mathbf{z}_{1:k}, \mathbf{\hat{z}}_{k+K}, \lambda_{k+K})) \\
    \approx \log \left( \sum_{i=1}^{N_{H}} p(\mathbf{\hat{z}}_{k+K} ~|~ \mathbf{\hat{x}}_{k+K}^{(i)})w_{k}^{(i)} \right) \\
    - \sum_{i=1}^{N_{H}}\log \left( p(\mathbf{\hat{z}}_{k+K} | \mathbf{\hat{x}}_{k+K}^{(i)}) \sum_{j=1}^{N_{H}} p(\mathbf{\hat{x}}_{k+K}^{(i)} | \mathbf{x}_{k}^{(j)})w_{k}^{(j)} \right)\hat{w}_{k+K}^{(i)}
\end{multline}
and otherwise, i.e., $\mathbf{\hat{z}}_{k+K} \not\in \mathcal{S} \cap \mathcal{O}^{c}$ at $k+K$,
\begin{multline}\label{eq:part_entropy_no_meas}
    H(p(\mathbf{\hat{x}}_{k+K}~|~\mathbf{z}_{1:k}, \mathbf{\hat{z}}_{k+K}, \lambda_{k+K})) \\
    \approx - \sum_{i=1}^{N_{H}}\log \left( \sum_{j=1}^{N_{H}} p(\mathbf{\hat{x}}_{k+K}^{(i)} ~|~ \mathbf{x}_{k}^{(j)})w_{k}^{(j)} \right)w_{k}^{(i)}
\end{multline}

EER takes the expectation of the information gain over all the possible measurements as shown in \eqref{eq:eer_def}. In order to reduce computation time, the number of particles to approximate the probability density function of $\mathbf{\hat{z}}_{k+K}$ is denoted by $N_{M}$ and is a user-defined parameter. In the expectation calculation, $N_{M}$ measurements are considered. Specifically, each of $N_{H}$ particles passes through the measurement model to consider $N_{M}$ possible measurements. In other words, from \eqref{eq:measurement_model}, each particle $\mathbf{x}_{k+K}^{(i)}$ for $i=1,...,N_{H}$, the expected measurements
\begin{equation}
    \mathbf{\hat{z}}_{k+K}^{(j)} = \textbf{h}(\mathbf{x}_{k+K}^{(i)}, \mathbf{n}_{k}),
\end{equation}
for $j=1,2,...,N_{M}$, are used to compute the expectation in \eqref{eq:eer_def}. Therefore, the EER is computed using the particle-based entropies derived before,
\begin{multline}
    EER(\lambda_{k+K}) \approx \mathbb{E}_{\hat{\mathbf{z}}_{k+K}} [ H(p(\mathbf{x}_{k}~|~\mathbf{z}_{1:k})) \\
    \quad - \sum_{j=1}^{N_{M}} H\left(p(\mathbf{\hat{x}}_{k+K}~|~\mathbf{z}_{1:k},\mathbf{\hat{z}}_{k+K},\lambda_{k+K})\right)
    p(\mathbf{\hat{z}}_{k+K}^{(j)} ~|~ \mathbf{\hat{x}}_{k+k}) ]  \label{eq:eer_particle}
\end{multline}

Then, the proposed guidance law computes the waypoint, or goal position, at $k$ such that the EER is maximized, 
\begin{equation}
\label{eq:maxEER}
    \lambda_{k} = \argmax_{\lambda_{k+K}\in\mathcal{A}_{k+K}} EER(\lambda_{k+K})
\end{equation}
where $\mathcal{A}_{k+K}$ is the set of possible waypoints for the agent at $k+K$.


\section{EXPERIMENT SETUP}
\label{sec:setup}

\begin{figure}[t]
    \centering
    \smallskip
    \smallskip
    \smallskip
    \includegraphics[width=0.4\textwidth]{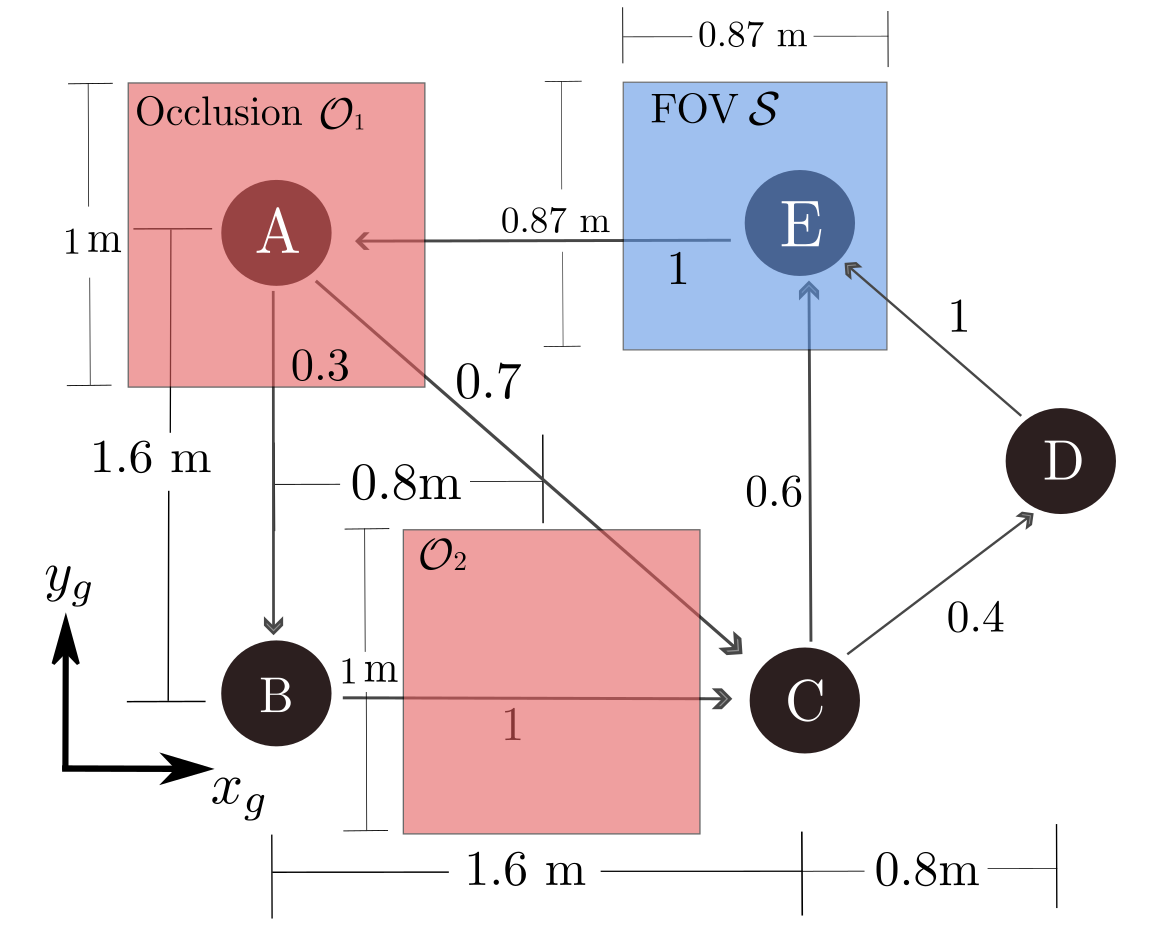}
    \caption{Markov chain-like road network model used in experiments.}
    \label{fig:road_network}
\end{figure}

The target TurtleBot follows a road network, which is modeled as a Markov chain. As shown in Figure \ref{fig:road_network} the target can move from one node to another based on the transitioning probability, denoted in the edges in the Figure. 
This Markov chain-like road network model is unknown to the agent, and therefore, learned using the DMMN (Section \ref{ssec:mml}). 
Figure \ref{fig:road_network} shows the dimensions of the road network and where the occlusion is located. Note that there are two occlusions, one between node B and C and one in node A where one of the two stochastic decisions takes place.

\begin{figure}[b]
    \centering
    \begin{subfigure}[t]{0.45\linewidth}
        \centering
        \includegraphics[width=3.5cm]{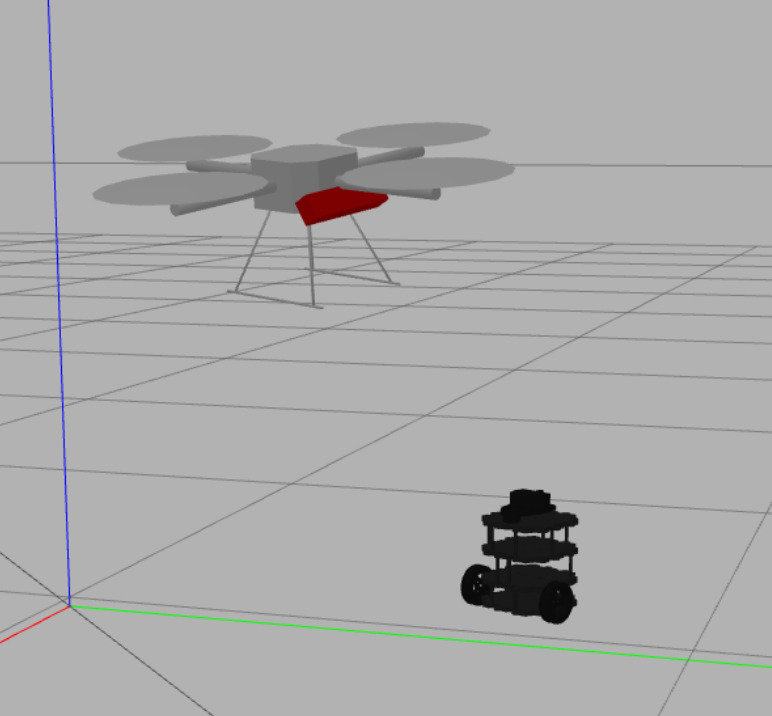}
        \caption{}
        \label{fig:gazebo_sim}
    \end{subfigure}%
    \hfill%
    \begin{subfigure}[t]{0.45\linewidth}
        \centering
        \includegraphics[width=3.7cm]{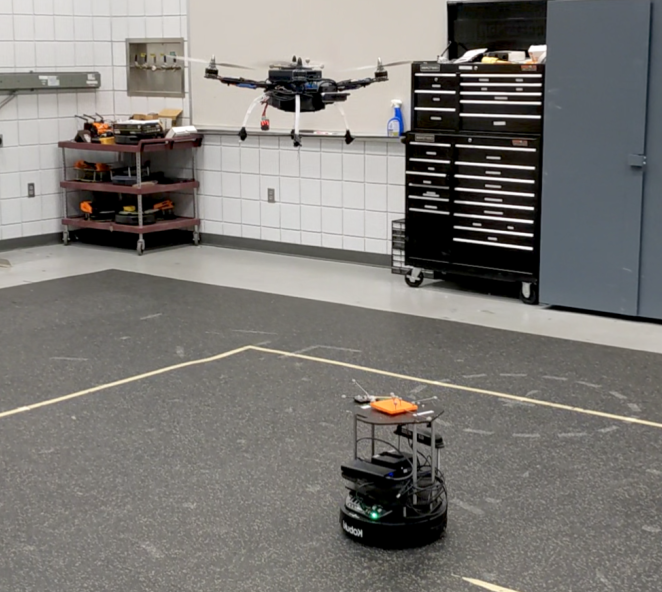}
        \caption{}
        \label{fig:hardware}
    \end{subfigure} 
    \caption{(a) Gazebo simulation and (b) hardware setup of the quadcopter agent and TurtleBot target.}
\end{figure}

This work is validated on hardware platforms at the Autonomous Vehicles Lab (AVL) located at the UF REEF, which houses a motion capture system with used to validate the true positions of the agent and target vehicles. The quadcopter is equipped with a NUC7i7 receiving the velocity commands from a computer running the particle filter along with the NN for predicting next target states, and the information-driven guidance. To be able to compute the filter with the DMMN, and the information-driven guidance in real time, they are run at 3 Hz and 2.5 Hz respectively. To localize itself and track the desired positions, the quadcopter computer runs the REEF Estimator and REEF Control \cite{ramos_reef_2019} as well as a DNN adaptive portion of the controller that reduces disturbances \cite{Lamb2023DeepNA}. The first results are 10 runs in simulation (Figure \ref{fig:gazebo_sim}) to compare estimation performance, and 3 runs in hardware (Figure \ref{fig:hardware})  to compare guidance performance.  
%
%
The hyper-parameters chosen in the simulation and in hardware are shown in Table \ref{tab:parameters}. We find that $K=4$ is appropriate for our setup because the future horizon depends on the ability of the agent to follow the goals and longer horizons resulted in the target being outside of FOV. 

\begin{table}[t]
    \smallskip
    \caption{Algorithm parameters used in simulation and hardware experiments}
    \centering
    \begin{tabular}{p{0.06\textwidth} p{0.3\textwidth} p{0.05\textwidth}}
    \hline
    Parameter    & Description                                    & Value\\
    \hline
    \hline
    $N$          & Number of samples used for prior approximation (particles)                     & $500$ \\
    $N_{H}$        & Number of samples used for entropy and EER computation   & $25$ \\
    $N_{M}$        & Number of future measurements considered in EER computation per sample & $1$ \\
    $K_{in}$     & Number target state histories input to transformer-based motion model  & $10$ \\
    $K$          & Time step horizon for waypoint planning       & $4$ \\
    $a$          & Multinomial resampling threshold         & $0.99$ \\
    $b$          & Uniform resampling threshold             & $0.4$ \\
    \hline
    \end{tabular}
    \label{tab:parameters}
\end{table}

\section{RESULTS}
\label{sec:results}

\subsection{Benchmark Comparison Setup}
\label{ssec:comparison_setup}
The novel tracking and guidance method is compared with two other guidance methods as baselines. The presented novel guidance law is referred to as DMMN-EER for convenience. The DMMN-EER method computes the next waypoint as the position of the maximum EER at the future $k+K$ time step as shown in \eqref{eq:maxEER}. 

The first baseline guidance method with which to compare the DMMN-EER is referred to as Lawnmower and Tracking (LAWN). LAWN computes the next waypoint as the position of the last target measurement, $\mathbf{z}_{k}$. If a target measurement is not available, LAWN outputs a lawnmower path (boustrophedon pattern) that covers the area of interest in $\mathcal{P}$.
This guidance law attempts to track the target without a target state estimation algorithm. 
The second baseline method is referred to as Particle Filter Weighted Mean (PFWM). PFWM computes the next waypoint as the estimated target position at $k$ using the weighted mean of the particle filter, denoted by
\begin{equation}
	\mu_{k} = \sum_{i = 1}^{N} w_k^{(i)} \mathbf{x}_k^{(i)}
\end{equation}

The third and final baseline method leverages a Kalman Filter (KF) by estimating the velocities of the target after feeding noisy position data. The guidance law for this baseline is similar to PFWM in which the agent's goal position is the mean position estimate of the filter. 

Four metrics are used to compare the performance among the four guidance methods described above. First, the tracking error, $e$, is defined by the $xy-$distance between the agent and target on $\mathcal{P}$, i.e.,
\begin{equation}
    e = || \mathbf{s}_{k} - \mathbf{x}_{k} ||_2
\end{equation}
Second, the estimation error, $\tilde{e}$, is defined by the distance between the the actual target position and the estimated target position computed by the weighted mean of the particle filter, i.e.,
\begin{equation}
\label{eq:e_estimation}
    \tilde{e} = || \sum_{i = 1}^{N} w_{k}^{(i)} \mathbf{x}_{k}^{(i)} - \mathbf{x}_{k} ||_{2}  
\end{equation}
Third, the determinant of the covariance matrix, $\det({\Sigma_k})$, is used to measure the target estimation uncertainty at time $k$. The covariance matrix of the particle states is computed by taking the sum of the difference between each particle and the weighted mean, i.e.,
\begin{equation}\label{eq:det_cov}
    \Sigma_{k} = \left( \sum_{i = 1}^{N} w_{k}^{(i)}(\mu_{k} - \mathbf{x}_k^{(i)}) \right) \left( \sum_{i = 1}^{N}w_{k}^{(i)}(\mu_{k} - \mathbf{x}_k^{(i)}) \right)^\top    
\end{equation}
The fourth metric, named $\%FOV$ is the percentage of the total time the target is seen by the agent's FOV and therefore there is a measurement, ie. $\mathbf{x} \in \mathcal{S} \cap \mathcal{O}^{c}$. 

\subsection{Motion Model Learning Results}
\label{ssec:mml_results}
The DMMN model, presented in Section \ref{ssec:mml}, is first implemented and tested in order to demonstrate that the motion model can accurately propagate the target state. The DMMN is trained offline from half an hour of noisy position data ($\sim3$k samples) collected in the the road network described in Figure \ref{fig:road_network}, which we assume we have access beforehand.

To demonstrate state estimation effectiveness, the DMMN is compared to two baseline motion models, where the first also uses a PF but estimates the target's next position using an estimated velocity computed by the change in observed position from when the target was in the FOV. This velocity is randomly sampled from a range of $[0, 0.7]$ m/s. The second baseline is a Kalman Filter (KF) that also estimates the target velocity when given noisy position inputs. The experiments consist of the identical duration (1.5 mins) and setup as the guidance experiments explained in Section \ref{exp_results} except the guidance goal is the true target position. The simulation 10 runs with the DMMN motion model results with a estimation error of $\tilde{e} = 0.361$ m with standard deviation of $0.281$ m, the particle velocity baseline motion model results in $\tilde{e} = 0.917$ m with standard deviation of $0.608$ m and the KF results in $\tilde{e} = 1.318$ m with standard deviation $1.145$ m. These results clearly shows that the DMMN helps achieving a lower error in estimation.

\subsection{Hardware Guidance Experiment Results}\label{exp_results}

\begin{figure}[t]
    \centering
    \includegraphics[width=0.47\textwidth]{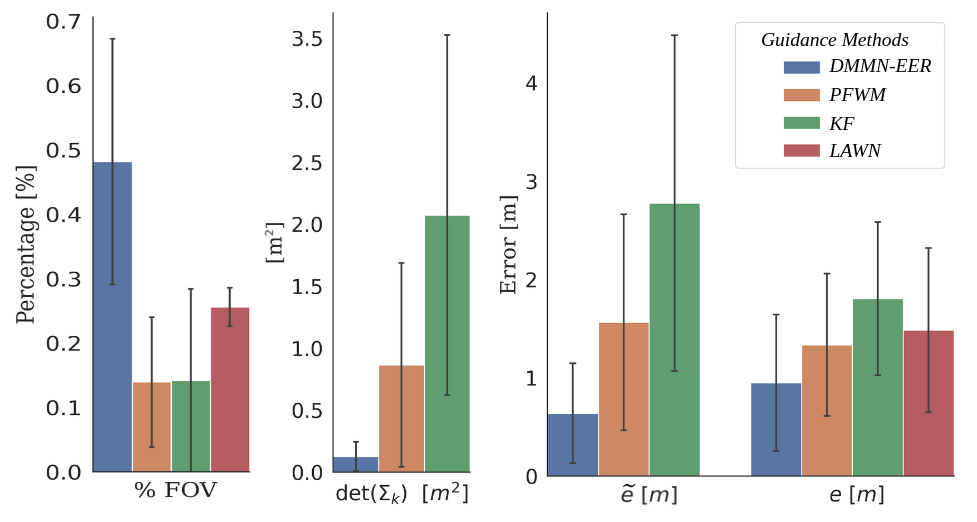}
    \caption{Hardware performance metrics comparison among the DMMN-EER (presented), LAWN, KF and PFWM (baselines) guidance methods.}
    \label{fig:bar_errors}
\end{figure}

For each guidance method, the identical experiment is again performed ten times to obtain average values of performance metrics. All four guidance methods use the same DMMN as an estimation method. In Figure \ref{fig:bar_errors}, the target tracking performance is compared using the four metrics defined in \ref{ssec:comparison_setup}. The figure shows the DMMN-EER method outperforms the other two baseline methods in each of the four quantitative evaluations metrics. When compared with the LAWN and PFWM methods, the DMMN-EER method achieves the smallest state and tracking error with the lowest distribution uncertainty and highest percent time in FOV.  

\begin{figure}[!ht]
    \centering
    \smallskip
    \begin{subfigure}[b]{0.35\textwidth}
        \centering
        \includegraphics[width=\linewidth]{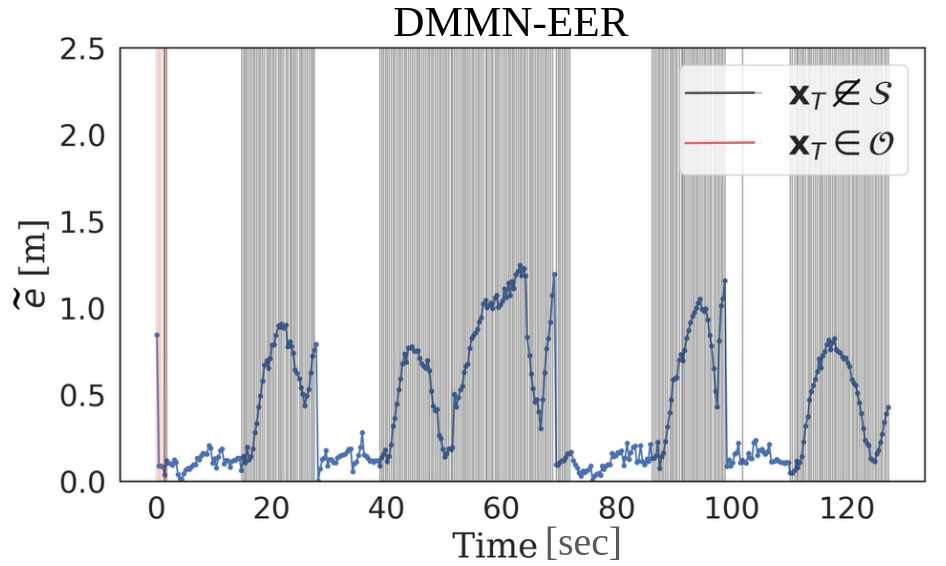}
    \end{subfigure}

    \vspace{5pt}  

    \begin{subfigure}[b]{0.35\textwidth}
        \centering
        \includegraphics[width=\linewidth]{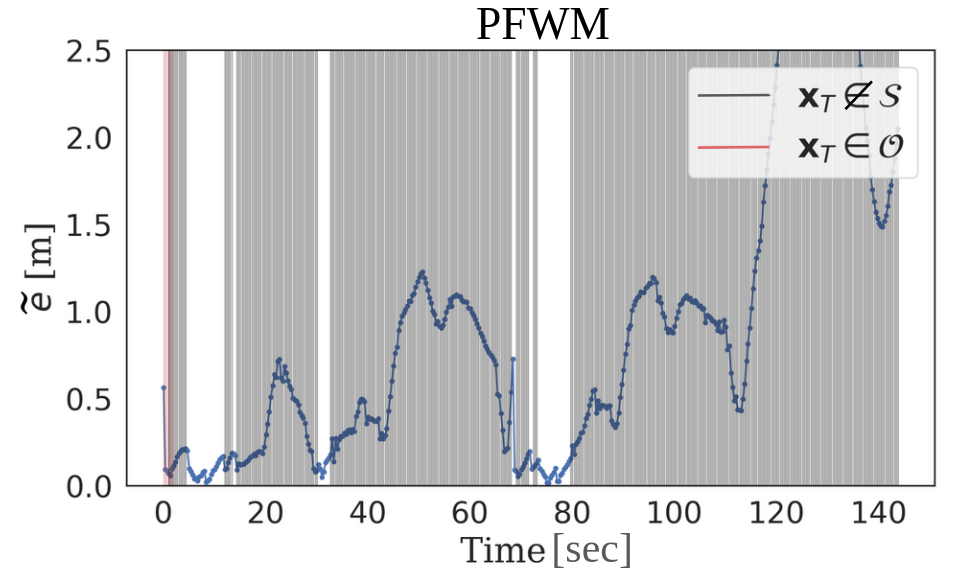}
    \end{subfigure}

    \begin{subfigure}[b]{0.35\textwidth}
        \centering
        \includegraphics[width=\linewidth]{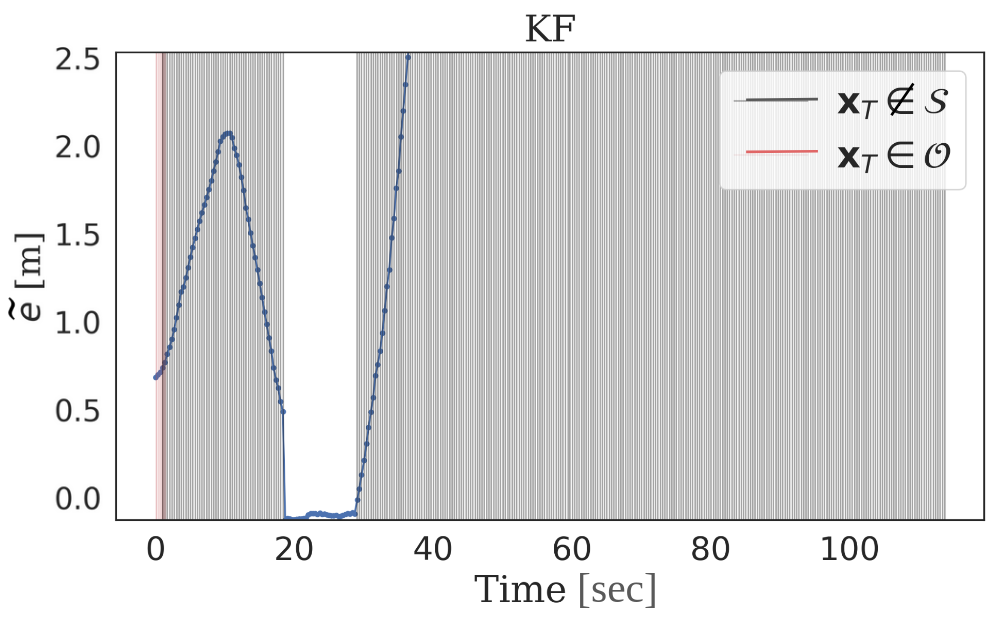}
    \end{subfigure}
    
    \begin{minipage}{\columnwidth}  
        \caption{Estimation error $(\tilde{e})$ over time for a single run of the DMMN-EER (presented), PFWM and KF (baselines) guidance methods from hardware experiments.}
        \label{fig:estimation_error_subfigures}
    \end{minipage}
\end{figure}

The advantage of DMMN-EER is highlighted in Figure \ref{fig:estimation_error_subfigures}. In the figure, the estimation error over time of a single run is plotted, where the red vertical lines are the times when the target is occluded, and black vertical lines are the times where the target is not inside the FOV. As shown in the figure, the DMMN-EER method has a lower estimation error over time, $\tilde{e}$, compared to the PFWM method. Moreover, it is noteworthy that the DMMN-EER can reduce $\tilde{e}$ more quickly than the other two when the target escapes the unknown occlusion zone. For all methods, the figure shows the error is increased inside and near the occlusion zones. While PFWM still has high $\tilde{e}$ right after the occlusion, DMMN-EER reduces $\tilde{e}$ efficiently because the EER-based approach can command the agent to move to the position that the target is most probable to be located at, based on the prediction from DMMN.

\section{CONCLUSION}
This paper presented a novel guidance law for target tracking applications where the target motion model is unknown and sensor measurements are intermittent. The target's motion model is modeled as an attention-based deep neural network and trained using previous measurements. Then, this trained deep motion model network is used in the prediction step of a particle filter estimating the target state. The information-driven guidance law calculates the next goal position for the agent to achieve the maximum expected entropy reduction on target state estimation.
Hardware experiments are conducted to compare the guidance method to other two baseline methods. The experiment results show that the presented novel guidance method reduces the target state estimation and tracking errors and estimation uncertainty.

For future work, the authors aim to validate the reliability of the result by performing a statistical difference test (e.g., one-way ANOVA). Additionally, we aim to integrate the learned motion model with an EKF to test the benefit of the PF as well as avoid the pre-training stage for model learning and perform online learning.
Ultimately, we aim to explore how this uncertainty-aware guidance approach can be extended to a multi-target tracking problem in a multi-agent system. 

\section*{ACKNOWLEDGMENT}
This work was supported by the Task Order Contract with the Air Force Research Laboratory, Munitions Directorate at Eglin AFB, AFOSR under Award FA8651-22-F-1052 and FA8651-23-1-0003. The authors would like to thank Jared Paquet for assisting in running the hardware experiments in the Autonomous Vehicle Lab (AVL) at the UF REEF and also Aditya Penurmati for providing the Kalman Filter baseline and running experiments at the APRILab at UF.

\bibliographystyle{IEEEtran}
\bibliography{mml}
\end{document}